\documentclass[twocolumn]{aastex631}

\usepackage{comment}
\usepackage{rotating}
\usepackage{ulem}
\usepackage{soul}
\usepackage{multirow}
\usepackage{longtable}
\usepackage{indentfirst}  
\usepackage{longtable}  
\usepackage{tablefootnote}
\shorttitle{Pulsars in M13}
\shortauthors{Yin et al.}

\graphicspath{{./}{figures/}}

\begin{document}

\title{Illuminating Hidden Pulsars: Scintillation-Enhanced Discovery of Two Binary Millisecond Pulsars in M13 with FAST}

\author[0000-0001-6051-3420]{Dejiang Yin}
\affiliation{College of Physics, Guizhou University, Guiyang 550025, China}

\author[0000-0003-0757-3584]{Lin Wang}
\affiliation{Shanghai Astronomical Observatory, Chinese Academy of Sciences, 80 Nandan Road, Shanghai 200030, China}
\affiliation{Kavli Institute for Astronomy and Astrophysics, Peking University, Beijing 100871, China}
\author[0000-0002-2394-9521]{Li-yun Zhang}
\affiliation{College of Physics, Guizhou University, Guiyang 550025, China}
\affiliation{Guizhou Radio Astronomical Observatory, Guizhou University Guiyang 550025, China}

%\affiliation{International Centre of Supernovae, Yunnan Key Laboratory, Kunming 650216, China}
% 
\author[0000-0003-0597-0957]{Lei Qian}
\affiliation{National Astronomical Observatories, Chinese Academy of Sciences, 20A Datun Road, Chaoyang District, Beijing 100101, China}
\affiliation{Guizhou Radio Astronomical Observatory, Guizhou University Guiyang 550025, China}
\affiliation{CAS Key Laboratory of FAST, National Astronomical Observatories, Chinese Academy of Sciences, Beijing 100101, China}
\affiliation{College of Astronomy and Space Sciences, University of Chinese Academy of Sciences, Beijing 100049, China}

\author{Baoda Li}\affiliation{College of Physics, Guizhou University, Guiyang 550025, China}

\author[0000-0002-2953-7376]{Kuo Liu}
\affiliation{Shanghai Astronomical Observatory, Chinese Academy of Sciences, 80 Nandan Road, Shanghai 200030, China}
\affiliation{State Key Laboratory of Radio Astronomy and Technology, A20 Datun Road, Chaoyang District, Beijing, 100101, China}

\author[0000-0001-6956-6553]{Bo Peng}
%\affiliation{National Astronomical Observatories, Chinese Academy of Sciences, Beijing 100101, China}
\affiliation{National Astronomical Observatories, Chinese Academy of Sciences, 20A Datun Road, Chaoyang District, Beijing 100101, China}
\author[0009-0007-6396-7891]{Yinfeng Dai}
\affiliation{College of Physics, Guizhou University, Guiyang 550025, China}

\author{Yaowei Li}
\affiliation{College of Physics, Guizhou University, Guiyang 550025, China}
% 
% 
% \author[0009-0001-6693-7555]{Yujie Lian} 
% \affiliation{Institute for Frontiers in Astronomy and Astrophysics, Beijing Normal University, Beijing 102206, China}
% \affiliation{Department of Astronomy, Beijing Normal University, Beijing 100875, China}
% 
% 
% \author[0000-0002-9322-9319]{Zhen Yan}
% \affiliation{Shanghai Astronomical Observatory, Chinese Academy of Sciences, Shanghai 200030, China}
% 
% 
\author[0000-0001-7771-2864]{Zhichen Pan}
\affiliation{National Astronomical Observatories, Chinese Academy of Sciences, 20A Datun Road, Chaoyang District, Beijing 100101, China}
\affiliation{Guizhou Radio Astronomical Observatory, Guizhou University Guiyang 550025, China}
\affiliation{CAS Key Laboratory of FAST, National Astronomical Observatories, Chinese Academy of Sciences, Beijing 100101, China}
\affiliation{College of Astronomy and Space Sciences, University of Chinese Academy of Sciences, Beijing 100049, China}

\correspondingauthor{}\email{wanglin@shao.ac.cn; liy\_zhang@hotmail.com, lqian@nao.cas.cn}

% \collaboration{20}{(AAS Journals Data Editors)}
% \author{3}
% \affiliation{1111}
% \affiliation{2222 \\
% 1667 K Street NW, Suite 800 \\
% Washington, DC 20006, USA}

\begin{abstract}
We conducted a sensitive acceleration search using Fast Fourier Transform (FFT) techniques on full-length and segmented data from 84 observations of the globular cluster M13 with the Five-hundred-meter Aperture Spherical radio Telescope (FAST). Employing a low detection threshold (2\,$\sigma$) to maximize sensitivity to faint pulsars, here we report the discovery of two binary millisecond pulsars: J1641+3627G (M13G) and J1641+3627H (M13H). Both pulsars were detected during scintillation-brightened states, revealing systems that would otherwise remain undetected. For M13G, we obtained a phase-connected timing solution spanning 6.4 years, identifying it as a black widow system with an orbital period of 0.12 days hosting an extremely low-mass companion ($\sim 9.9\times 10^{-3}~{ M}_\odot$), though no eclipses were observed. M13H, however, shows significant apparent acceleration but was detected in only 2 of 84 observations; its extremely low detection rate currently prevents constraints on orbital parameters or classification. 
\end{abstract}


\section{Introduction}
\label{Intro}
% GC psr 
Globular Clusters (GCs) are compact and massive stellar systems containing $\sim10^4-10^6$ stars, with stellar densities of more than $\sim10^3~\rm M_{\odot}\rm pc^{-3}$ (\citealt{1996AJ....112.1487H}, 2010 edition).
Exotic pulsars in GCs often originate from frequent dynamical interactions between compact objects within these dense stellar environments \citep{2006MNRAS.372.1043I}.
%\textcolor{blue}{\sout{Thus}}
Pulsars in GCs can serve as effective probes for investigating the cluster dynamics (e.g., \citealt{1993ASPC...50..141P}).
Since \citet{1987Natur.328..399L} discovered the first GC pulsar in M28, a total number of 345 radio pulsars have been detected in 45 GCs\footnote{\url{https://www3.mpifr-bonn.mpg.de/staff/pfreire/GCpsr.html}}.
Future detections of GC pulsars are expected to increase significantly with the enhanced sensitivity of next-generation telescopes, such as the Square Kilometre Array (SKA) \citep{2015aska.confE..47H}, as well as through advanced search techniques, including jerk-search algorithms that accounts for higher-order orbital dynamics \citep{2018ApJ...863L..13A}.

Monitoring and studying GC pulsars enable diverse scientific goals.
It allows us to obtain or update timing solutions for known pulsars (e.g., \citealt{2017MNRAS.471..857F}), and to measure additional astrometric parameters such as proper motion, which are essential for studying cluster dynamics (e.g., \citealt{2017MNRAS.471..857F}).
Moreover, precise timing can lead to the measurement of post-Keplerian parameters, offering constraints on the masses of pulsar systems (e.g., \citealt{2016ARA&A..54..401O}), as well as enables detailed studies of timing noise (e.g., \citealt{2024arXiv241021648R}).
%The discovery of new pulsars is not direct product of long-term GC pulsar timing campaigns.
Additionally, for pulsars whose signals are strongly modulated by interstellar scintillation, resulting in low detection rates, extended and repeated monitoring campaigns are essential to ensure their detection (e.g., \citealt{2016MNRAS.459L..26P,2016MNRAS.462.2918R}).
Especially pulsars with relatively low DM (e.g., DM $\lesssim$ 50) often appear significant modulation of apparent flux densities in both time and frequency due to the stellar scintillation effects \citep{1990ARA&A..28..561R, 2023PASA...40...21B}.
For example, the six known pulsars A to F in M13 have exhibited significant variations in flux density, and the non-detections of those pulsars in any epochs are most likely attributable to interstellar scintillation effects \citep{2020ApJ...892...43W}.
A screen-like scattering medium was detected at distances 6.7$^{+0.2}_{-0.2}$ kpc in the line-of-sight direction of M13, and the scattering screens are located at 4.4$^{+0.1}_{-0.1}$ kpc above the Galactic plane \citep{2023SCPMA..6699511Z}.
The scintillation timescale of M13A was in the range of 10.3$\pm$1.4 -- 23.4$\pm$2.5 minutes \citep{2023SCPMA..6699511Z}.
Thus, the flux densities of potential new pulsars in M13 can occasionally brighten due to scintillation effects, providing an opportunity to detect previously missed pulsars or those that may have been below the sensitivity threshold of the survey.

M13 (NGC~6205), also known as the Hercules GC, is located in the Milky Way halo at $\alpha_{\rm J2000} = 16^{\rm h}41^{\rm m}41^{\rm s}.24$, $\delta_{\rm J2000} = +36^{\circ}27'35''.5$. Its core and half-light radii are $r_{\rm c} = 0.62'$ and $r_{\rm h} = 1.67'$, respectively \citep[][2010 edition]{1996AJ....112.1487H}, almost within the $L$-band half-power beamwidth (2.9$'$) of the Five-hundred-meter Aperture Spherical radio Telescope (FAST; \citealt{2011IJMPD..20..989N}; \citealt{2020Innov...100053Q}).
Prior to this work, seven pulsars had been reported in M13 (M13A-B, \citealt{1993PhDT.......136A}; M13C-E, \citealt{2007ApJ...670..363H}; M13F, \citealt{2020ApJ...892...43W}; M13I, \citealt{2025arXiv250505021L}).
M13 was previously estimated to host about 5 detectable pulsars due to its relatively low stellar encounter rate \citep{2013MNRAS.436.3720T}, while it was expected to host more pulsars in recent GC population simulation \citep{2024ApJ...969L...7Y}. 

In this work, we conducted deep searches for new pulsars using a total of 84 FAST observations of M13, spanning seven years (2018 -- 2025), which had been carried out to conduct a long-term  timing of previously known pulsars (Wang et al., in preparation). 
This number of observations increases the possibility of detecting potential 
weak pulsars that only rarely become detectable because of strong amplification by scintillation.
% pulsars with significant interstellar scintillation effect.
In order to detect new pulsars in this cluster,
we performed both a fast-folding algorithm search in the time domain (e.g., \citealt{2020MNRAS.497.4654M}) and an acceleration search in the frequency domain  (e.g., \citealt{2002AJ....124.1788R}) in parallel. This paper reports the results of the acceleration search, while the results of another search scheme are presented separately \citep{2025arXiv250505021L}.

The structure of this paper is as follows:
The FAST observations and data reduction scheme are presented in Section \ref{sec:Obs-Data}.
The new discoveries of pulsars and their timing results are given in Section \ref{sec:results}.
%\textcolor{blue}{\sout{The analysis of pulsar acceleration in}} 
%\textcolor{blue}{\sout{the cluster field is presented in Appendix \ref{GC_field}}}.
A brief discussion and summary are given in Section \ref{discussion} and Section \ref{Conclusions}.

\section{Observation and data reduction} \label{sec:Obs-Data}
\subsection{FAST Observation} \label{sec:floats}

As a priority target in the early stage of the FAST GC pulsar search, M13 has been observed by FAST since 2017, leading to the discovery of a new pulsar binary, M13F \citep{2020ApJ...892...43W}.
To obtain or update timing solutions for all known pulsars in M13, a total of 25 observations (2017 December 20 to 2019 September 14) were conducted, including one observation with the Ultra-Wide-Band receiver and 24 observations with the 19-beam receiver (see details in \citealt{2020ApJ...892...43W}).
After this stage, follow-up observations of this cluster were made approximately once a month to perform a high-precision timing analysis of the known pulsars.
The data were taken using the FAST 19-beam receiver, covering a frequency range of 1.0 -- 1.5\,GHz and channelized into 4096 channels, each with a width of 0.122\,MHz \citep{2019SCPMA..6259502J}.
All observations were performed with the \texttt{Tracking} mode or \texttt{SwiftCalibration} mode with a sampling time of $49.152\,\mu{\rm s}$ and packaged in the standard search-mode PSRFITS data format \citep{2004PASA...21..302H}.
During each observation, the central beam was pointed at the center of M13.

Most follow-up observations had a duration of approximately 4000 or 5000\,s.
To search for pulsars far away from the cluster center, a 8000\,s \texttt{SnapShot} mode observation was conducted on 2021 October 1, using a grid of four adjacent 2000\,s pointings to uniformly cover a wider region of diameter $\gtrsim 30^{\rm '}$.
Additionally, a long-duration tracking observation of 5\,hr was carried out (on 2023 November 8) to deeply search for faint pulsars.
In this study, only data from the central beam were analyzed for each observation.
We performed a pulsar search on 84 datasets, which span seven years, up to 2025 February 25.
The only archival data taken with the Ultra-Wide-Band receiver on 2017 December 20 was not used in this work.  

\subsection{Previous Pulsar searches for GC M13} \label{sec:searches}
In the early phase of FAST observations of M13, pulsar searches were conducted using acceleration search methods implemented in the PRESTO toolkit \citep{2001PhDT.......123R}, which led to the discovery of M13F; see \citet{2020ApJ...892...43W} for details.
We recently applied two parallel pipelines to search the FAST data of M13: one using an Fast Fourier Transform (FFT)-based \textsc{PRESTO} acceleration search and the other employing a Fast Folding Algorithm (FFA)-based pipeline implemented with the \textsc{Riptide} package \citep{2020MNRAS.497.4654M}. 
The FFA-based pipeline resulted in the discovery of M13I in a moderate orbit of approximately 18 days; more details of this search scheme are provided in \citet{2025arXiv250505021L}.
The two new pulsars reported here, M13G and M13H, were missed in the earlier FFT-based search, and M13H was also missed in the recent FFA-based search. These non-detections are primarily due to their intrinsically weak signals and binary motion, which caused their detection significance to fall below the sifting thresholds in the earlier FFT-based search; in the case of M13H, its large apparent acceleration further limited its detectability in FFT-based methods.

\subsection{Data Reduction} \label{sec:floats}
A typical set of steps from the \textsc{PulsaR Exploration and Search TOolkit} (\textsc{PRESTO}\footnote{\url{https://github.com/scottransom/presto}},
\citealt{2001PhDT.......123R, 2002AJ....124.1788R}) search suite was applied to find new pulsars.
The \texttt{rfifind} routine in \textsc{PRESTO} was used to perform the mitigation of radio frequency interference (RFI) in both the time and frequency domains and a mask file was generated at this stage.
The dedispersed time series over a range of trial dispersion measure (DM) values were generated from PSRFITS data using the \texttt{prepsubband} routine.
The DMs of previously known pulsars (A to F) in M13 range from approximately 29.45~cm$^{-3}$\,pc (M13B)  to 31.28~cm$^{-3}$\,pc (M13E), with an average of 30.23~cm$^{-3}$\,pc.
The DM range for potential new pulsars in M13 was estimated to be 29.18~cm$^{-3}$\,pc to 31.28~cm$^{-3}$\,pc, based on the empirical power-law relationship between the average DM and DM spread of known pulsars in GCs \citep{2023RAA....23e5012Y}.
We ultimately selected DM trials ranging from 28~cm$^{-3}$\,pc to 32~cm$^{-3}$\,pc, with a step size of 0.05~cm$^{-3}$\,pc.
This leads to a pulse broadening of only $\sim$2$\%$ compared to the shortest spin period of 0.5~ms considered in our search (e.g., \citealt{2007ApJ...670..363H}).

The dedispersed time series were transformed into the fluctuation frequency domain for subsequent Fourier domain acceleration searches using the \texttt{realfft} routine.
Low-frequency red noise in the frequency domain was removed using the \texttt{rednoise} routine.
The \texttt{accelsearch} routine was then employed to perform Fourier domain acceleration searches, utilizing harmonic summing of 32 to account for pulses with a narrow duty cycle.
The \texttt{$z_{\rm max}$} flag in \texttt{accelsearch} corresponds to the number of Fourier frequency bins that the signal drifts through over the course of the observation.
This parameter is used to correct for Doppler smearing by assuming approximately constant acceleration, defined as
$a_{\rm l}hfT^{2}_{\rm obs} / c$, where $a_{\rm l}$ presents the constant acceleration caused by underlying orbital motion, $h$ is the harmonic summing factor, $T_{\rm obs}$ is the observation duration, f is the fundamental frequency and $c$ denotes the speed of light.
We searched all full-length data for isolated pulsars using \texttt{$z_{\rm max}$} = 20. 
%\textcolor{blue}{\sout{This value is chosen to account for the expected}} 
%\textcolor{blue}{\sout{ line-of-sight accelerations induced by the cluster's gravitational}}  \textcolor{blue}{\sout{potential (on the order of 10$^{-9}\rm \,m\,s^{-2}$ for M13; \citealt{1993ASPC...50..141P}).}}
We conducted searches on both segmented data and full-length data for accelerated pulsars in compact orbits using \texttt{$z_{\rm max}$} values of 300 and 520, respectively. This acceleration search was sensitive to accelerations up to $\pm$195.0~$\rm m~s^{-2}$ for a 200~Hz signal (assuming $h = 1$, i.e., the fundamental harmonic) during a 2000~s observation.

The effectiveness of the acceleration search technique depends on the ratio of the integration length of observation to the pulsar's orbital period (e.g., \citealt{2002AJ....124.1788R}) and on the orbital phase at which the pulsar is observed (e.g., \citealt{2013MNRAS.431..292E, 2015MNRAS.450.2922N}).
To search for accelerating pulsars in short orbits, observations are typically divided into multiple sub-integration segments. Each halving of the integration time reduces the minimum detectable orbital period by half, but results in a $\sqrt{2}$ decrease in the sensitivity of the flux density.
In addition, pulsars in M13 may exhibit short-timescale modulation of pulse intensity on the order of minutes due to diffraction scintillation (e.g., \citealt{2023SCPMA..6699511Z}).
To balance these trade-offs, we employed a so-called \textbf{\textit{MOSS}\footnote{\textsc{Multiple Observation Segment Search (\textit{MOSS}) for Pulsars:} \url{https://github.com/ydejiang/MOSS}}} script to evenly divide the observational data into segments of equal duration, allowing for arbitrary time overlaps between adjacent segments to enhance search sensitivity.
We refer to this scheme as the ``overlapping coherent segmented acceleration search". 
This scheme increases the probability of blindly detecting pulsars in short-orbit systems, which may simultaneously exhibit scintillation and/or likely eclipses phenomena.

Taking into account the duration of observation across different epochs, all data were evenly divided into approximately 1500 to 2500~s segments for acceleration searches.
For the longest 18,075 s tracking observation on November 8, 2023, we evenly divided the total observation time into segments of different durations, including 400\,s, 1000\,s, 2250\,s and 4000\,s, employing overlapping strategies of 0\,$\%$, 20\,$\%$ and 50\,$\%$ between adjacent segments to search with same parameters.
Finally, the sifting code  \texttt{JinglePulsar\footnote{\url{https://github.com/jinglepulsar/jinglesifting}}} \citep{2021RAA....21..143P} was applied to filter all candidates obtained from the \texttt{accelsearch} routine.
The selected candidates were then folded using \textsc{PRESTO}’s \texttt{prepfold} routine to generate standard diagnostic plots for visual inspection.

\section{Results} \label{sec:results}

\subsection{new discoveries}

Our acceleration search scheme successfully identified two millisecond pulsar binaries in M13, PSRs J1641+3627G and J1641+3627H (designated M13G and M13H). The first detection of M13G occurred during the acceleration search analysis of our longest tracking observation on 2023 November 8. For this observation, we divided the data into eight equal segments of 2250\,s each, % \textcolor{red}{
yielding a total integration time of 18000\,s.
M13G exhibits a spin period of 4.32\,ms and a DM of 30.75\,cm$^{-3}$\,pc. 
%\textcolor{blue}{\sout{M13G exhibits a complex average pulse profile}} \textcolor{blue}{\sout{characterized by several distinct components.The main pulse }}
%\textcolor{blue}{\sout{consists of two peaks, with the leading component being }}
%\textcolor{blue}{\sout{slightly brighter than the trailing one. An interpulse }}
%\textcolor{blue}{\sout{appears at a phase offset of 0.5 from the main pulse,}}
%\textcolor{blue}{\sout{while additional pulse locating}}
%\textcolor{blue}{\sout{between the main}}
%\textcolor{blue}{\sout{pulse and interpulse.}}
It exhibits a complex pulse profile, see Figure 1 and detailed description in Appendix~A.
We folded the full observation using the preliminary ephemeris from discovering epoch, covering 1.7 orbits, and found no significant short-timescale scintillation or eclipse.

M13H (see Figure \ref{fig:M13H}) was first detected in the initial 2000\,s segment of the \texttt{SnapShot} mode observation on 2021 October 1.
The optimal DM for this pulsar during this observation was 30.75\,cm$^{-3}$\,pc, with apparent barycentric spin period of 11.208571(4)\,ms and a significant acceleration of $-$14.42(19)\,m\,s$^{-2}$.
M13H was confirmed in the second approximately 2500\,s segment of the observation on January 23, 2023.
In the second detection, the DM, barycentric spin period, and acceleration of M13H were 30.80 $\mathrm{cm}^{-3}$\,pc, 11.213921(3)\,ms, and $-$24.60(12)\,$\mathrm{m\ s}^{-2}$, respectively.
%\textcolor{blue}{\sout{The apparent spin period derivative of M13H}}
%\textcolor{blue}{\sout{is on the order of 10$^{-11}$\,s\,s$^{-1}$.}}
The pulse profile of M13H is also described in detail in Appendix~A.

We then conducted full-length acceleration searches on the confirmation observation using \texttt{$z_{\rm max}$} of 300, 500 and 1200. While M13H remained detectable in the full-length search, the significance was slightly reduced compared to the segmented searches, yielding values of 3.8\,$\sigma$ and 4.0\,$\sigma$, respectively. No significant scintillation was observed during the 5000\,s confirmation observation, suggesting that the reduced significance is more likely attributable to orbital acceleration effects. 

\begin{figure*}[htp!]
\begin{center}
\includegraphics[width=0.9\linewidth]{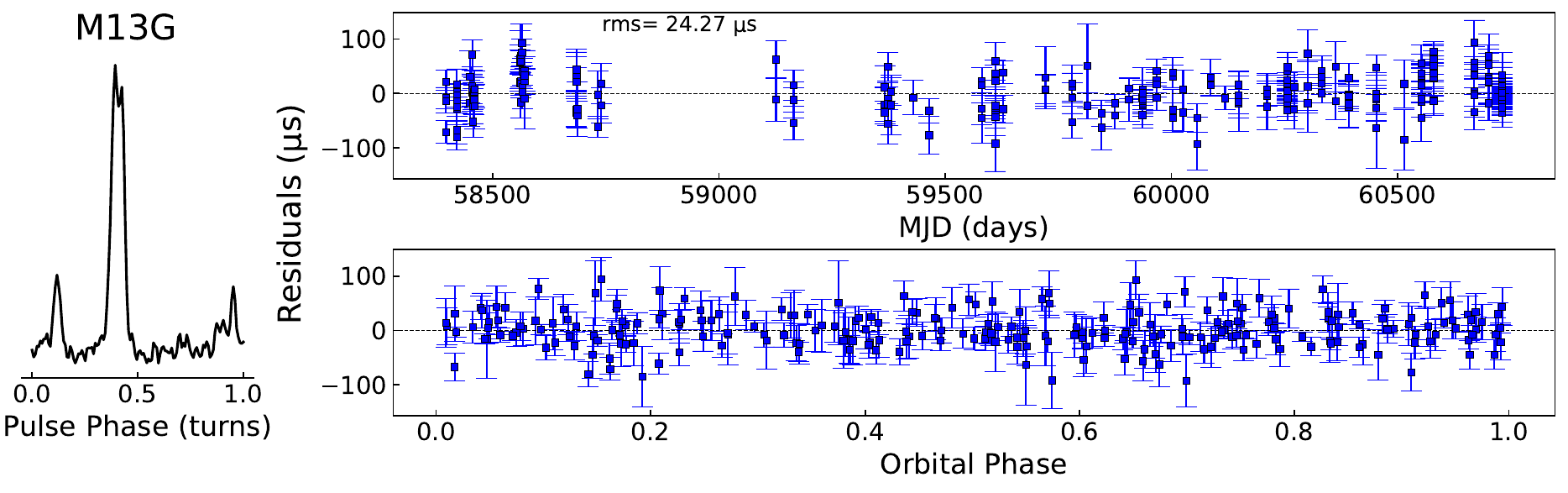}
\caption{
The average pulse profile and timing residual of M13G. The left panel is the integrated pulse profile obtained by summing all 57 detections over 128 pulse phase bins.
The timing residual as a function of MJD (upper subplot) and orbital phase (lower subplot) from the best-fit timing model are shown in the right panel.
}\label{fig:timing}
\end{center}
\end{figure*}

\subsection{The timing of M13G}

The \texttt{fitorb.py} routine from \textsc{PRESTO} was used to derive its initial orbital parameters based on the apparent barycentric spin periods and their derivatives of a pulsar at different orbit phases. We then folded the data using spin parameters from searching campaign with the \texttt{prepfold} routine. 
We created a standard profile for M13G by fitting Gaussian components to the discovery data (2023 November 8) using \textsc{PRESTO}'s \texttt{pygaussfit.py}.
Time-of-arrivals (TOAs) were obtained using \texttt{get\_TOA.py} routine of PRESTO  based on the Fourier-domain template matching described by \citet{1992RSPTA.341..117T}. %\textcolor{red}{how do you obtain a template? e.g. which dataset, is it noise free ...?}
For the timing analysis, we employed both the \textsc{TEMPO}\footnote{\url{https://tempo.sourceforge.net/}} \citep{2015ascl.soft09002N} and \textsc{Dracula}\footnote{\url{https://github.com/pfreire163/Dracula}} \citep{2018MNRAS.476.4794F} software packages to obtain a phase-connected timing solution.

The initial ephemeris was determined from our longest and most sensitive observation, obtained on 2023 November 8. 
We iteratively refined the folding and timing procedures until achieving a final solution with a reduced $\chi^2 < 2$, ensuring a robust timing model.
The phase-connected timing solution for M13G is presented in Table \ref{table:timing}, with the corresponding timing residuals shown in Figure \ref{fig:timing}. 
The mass function is given by (e.g., \citealt{2016ARA&A..54..401O}):
\begin{equation} \label{eqn:mass_function}
    f(M_\mathrm{p},M_\mathrm{c}) = \frac{(M_\mathrm{c} \sin i)^3}{(M_\mathrm{p} + M_\mathrm{c})^2} = \frac{4\pi^2}{T_{\odot}}\frac{x_{\rm psr}^3}{P_\mathrm{b}^2},
\end{equation}
where $T_{\odot} = GM_{\odot}/c^3 = 4.925490947\,\mu \rm s$ is the mass of the Sun in time units (the Newton’s gravitational constant $G$, the speed of light $c$ and the solar mass $M_{\odot}$). $i$ is the orbital inclination angle of system, $M_{\rm p}$ and $M_{\rm c}$ are the masses of pulsar and its companion.
Assuming $M_\mathrm{p}\, = \, 1.35 \, M_{\odot}$, we obtained a minimum ($i\, = \,90^\circ$) and median ($i\, = \,60^\circ$) companion masses of 8.6$\times 10^{-3}$ and 9.9$\times 10^{-3}$  $M_\odot$ respectively.
This suggest that M13G belongs to the black widow class.  %%%%%
The gravitational potential from its host cluster M13 can account for its observed negative spin period derivative (see Appendix \ref{GC_field} for details).

\section{discussion}\label{discussion} 
\subsection{Properties of new discoveries}

Among all known pulsars in M13, M13G exhibits the second-largest projected angular offset from the cluster center at 1.8\,arcmin. This places it beyond both the cluster’s half-light radius (1.5\,arcmin) and the typical field of view of FAST observations. The resulting positional offset reduces the received flux density of M13G, and combined with potential stellar scintillation effects, likely contributed to its non-detection in certain observational epochs.
Using the timing-derived position of M13G, we searched for potential counterparts in archival observations from the Chandra  $X$-ray Observatory (0.3–8\,keV, ObsID:7290) and Hubble Space Telescope (optical/$UV$ bands; GO-10775, PI:Sarajedini and GO-12605, PI:Piotto). No significant counterpart was detected in either wavelength regime, with 3-$\sigma$ upper limits.  

\begin{figure}[h]
\begin{center}
\includegraphics[width=1\linewidth]{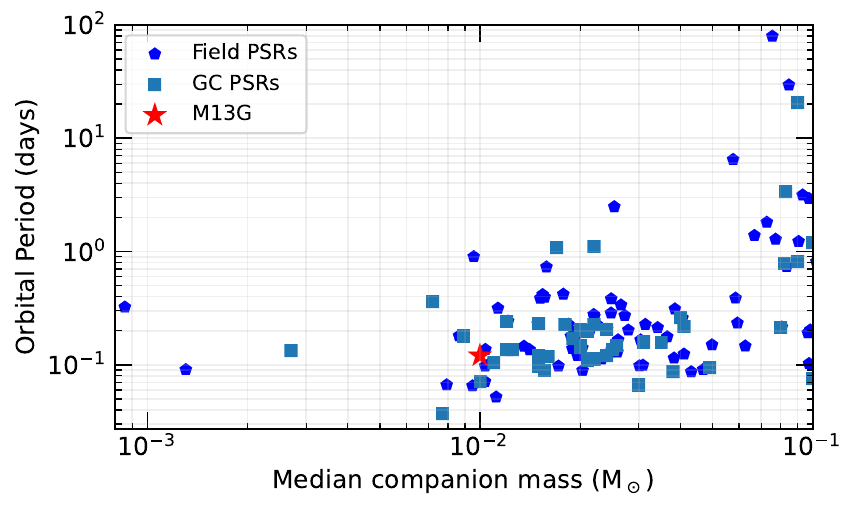}
\caption{
The median companion mass vs. orbital period (zoomed in on the range of $M_{\rm c, med}$ = $8\times 10^{-4}$ - $1\times10^{-1}$ ${\rm M}_\odot$.
The data was taken from the ATNF pulsar catalogue \citep{2005AJ....129.1993M} and GC pulsar catalog (As of 2025 February 22) with the \texttt{psrqpy} package \citep{2018JOSS....3..538P}.}
\label{fig:Mc-Pb}
\end{center}
\end{figure}

The companion star of M13G has an extremely low mass, ranking it as the sixth lowest-mass companion among GC pulsars (Figure \ref{fig:Mc-Pb}). The five GC pulsars with lower-mass companions include: M62H (0.0027\,${ M}_\odot$; potentially a planetary-mass companion, either a degenerate helium star or primordial planet, \citealt{2024MNRAS.530.1436V}), NGC~6440H (0.0072\,${ M}_\odot$; possibly an ultra-low-mass carbon white dwarf or brown dwarf, \citealt{2022MNRAS.513.1386V}), M71E (nature uncertain, may be a stripped dwarf or degenerate companion,  \citealt{2023Natur.620..961P, 2023ApJ...956L..39Y}), along with two confirmed black widow systems, 47~Tuc ac and NGC~6544A (though their companion types remain uncharacterized, \citealt{2021MNRAS.504.1407R, 2001ApJ...548L.171D}).

% plt.xlim(8e-4,1e-1)

No significant eccentricity was measured for M13G suggesting this pulsar is in a circular orbit, which is consistent with other black widow pulsars (e.g., \citealt{2021MNRAS.504.1407R}).
Binaries located far from the cluster center experience fewer stellar encounters compared to core systems, thereby preserving their circular orbits (e.g., \citealt{1996MNRAS.282.1064H}). 
The circular orbit is more consistent with tidal circularization processes \citep{1992RSPTA.341...39P}. No black widow systems in GCs have shown appreciable eccentricity, irrespective of the local stellar density; this could be in part due to their compact nature, which reduces the cross section for direct impacts.
%\textcolor{blue}{\sout{Besides, the circular orbit is also consistent }} \textcolor{blue}{\sout{with tidal circularization processes \citep{1992RSPTA.341...39P}.}} 
While this mechanism is most efficient for systems with low-mass He companions, the principle generally applies to close binaries with degenerate companions \citep{1992RSPTA.341...39P}.

No eclipse events were detected in M13G. This absence of eclipses is consistent with the system's low mass function, as spider binaries with higher inclination angles typically exhibit both eclipses and larger mass functions \citep{2019A&A...629A..92G, 2025A&A...698A.239B}. The observed properties suggest M13G likely has a relatively face-on orbital configuration, similar to other non-eclipsing black widow systems studied by \citet{2005ASPC..328..405F} and \citet{2025ApJ...979..143K}.

To determine whether M13H is located within the core region of M13, we searched the \texttt{snapshot} observation \citep{2020RAA....20...64J, 2021RAA....21..107H} conducted on October 1, 2021, covering beams M01 to M07 over a region of 4.5 arcmin from the center of M13. The signal from M13H was only detected in the beam pointing toward the GC center, suggesting that M13H is located in the core region of M13.
Additionally, following the assumption that the observed integration time ($T_{\rm obs}$) is a fraction of the orbital period ($T_{\rm orb}$) in the acceleration search algorithm ($T_{\rm obs}\lesssim P_{\rm orb}/10$), we estimated the orbital period of M13H to be at least 0.5 days. If we can detect this pulsar in subsequent monitoring observations, it will help determine its orbital parameters.

\subsection{The flux estimation of new pulsars}\label{flux}
We estimated flux densities in each detections of M13G and H with the radiometer equation \ref{eq:fluxestimate} \citep[see, e.g.,][]{1985ApJ...294L..25D}. 
\begin{equation}
S_{\rm min} = \beta \displaystyle  \frac{(S/N_{\rm min}) T_{\rm sys}}{G \sqrt{n_{\rm p} T_{\rm obs} \Delta f}}\sqrt{\frac{W_{\rm obs}}{P-W_{\rm obs}}},
\label{eq:fluxestimate}
\end{equation}
where $\beta \sim 1$ is the sampling efficiency for our 8-bit recording system, the $(S/N)_{\rm min}$ is the detected signal-to-noise ratio of pulsars, the system temperature ($T_{\rm sys}$) is $\sim$20~K, the antenna gain ($G$) is 16\,K\,Jy$^{-1}$, and the number of polarizations ($n_{\rm p}$) is 2, 
$T_{\rm obs}$ is the integration time and $\Delta f$ is the bandwidth in MHz, taken to be 400 MHz here from 1.05\,GHz to 1.45\,GHz \citep{2020RAA....20...64J}.

The estimated flux density range for M13G spans from 0.96 $\mu {\rm Jy}$ to 8.39 $\mu {\rm Jy}$, with a median value of 2.40 $\mu {\rm Jy}$.
The significant variation in the flux density of M13G (see Figure~\ref{fig:snr}) was possibly caused by the interstellar scintillation, like those of other known pulsars within this cluster \citep{2020ApJ...892...43W,2025arXiv250505021L}.

M13H was detected in only two of the 84 datasets mentioned above, despite attempts with  the ``jerk" search technique \citep{2018ApJ...863L..13A}.
The rough flux density range of our detections was 2.12 -- 3.48\,$\mu {\rm Jy}$ for M13H in these two epochs.
The main reason for the discovery of M13H is that interstellar scintillation fortuitously enhanced its flux density on the two observations. Thus, its intrinsic signal strength is probably even weaker.
The high acceleration and stellar scintillation make blindly searching for this pulsar difficult.

To detect more highly accelerated pulsars in M13, the ``jerk'' search technique was also used to search all the segmented data.
The acceleration search relies on a linearly-varying spin frequency for approximating constant acceleration, the jerk search depends on linearly-varying acceleration ($\dot{a_{\rm l}}$) or a quadratically-varying spin frequency ($\ddot{f}$) to approximate constant jerk.
The search of jerk technique was denoted by the $w_{\rm max}$ flag in \texttt{accelsearch} ($w_{\rm max}$ = $\dot{a_{\rm l}}hfT^{3}_{\rm obs} / c$, details see \citealt{2018ApJ...863L..13A}).
Due to the limited computing resources, we only searched the segmented data, using \(z_{\rm max} = 100\), \(w_{\rm max}=500\)  and \(z_{\rm max} = 300\), \(w_{\rm max}=900\) with a four-harmonic summing. No additional signals were identified during the ``jerk'' search process.

To date, nine radio pulsars (M13A–I) have been detected in M13. Previous studies predicted a range of population sizes in M13, with estimates of 25$\pm$11 \citep{2011MNRAS.418..477B}, $\sim$5 \citep{2013MNRAS.436.3720T}, and $\sim$24 \citep{2024ApJ...969L...7Y}.
The number of known pulsars in M13 is generally consistent with these predictions. If additional pulsars remain undetected in the FAST archival data, their non-detection is likely due primarily to effects of orbital acceleration in binary systems.
More sophisticated, yet computationally demanding, algorithms are expected to uncover highly accelerated pulsars in compact binaries (e.g., phase-modulation search, \citealt{2003ApJ...589..911R}; template-bank algorithm, \citealt{2022MNRAS.511.1265B}).

\subsection{The detection significance of weak pulsars}

Weak pulsars are expected to dominate the GC pulsar population under lognormal or power-law luminosity distributions, particularly for recycled pulsars (e.g., \citealt{2011MNRAS.418..477B, 2013MNRAS.431..874C}). Thus, although most weak pulsars lie below the survey’s sensitivity threshold, some with flux densities near the detection limit can still be identified through careful candidate selection and further inspection.
In \textsc{PRESTO} search, the sigma ($\sigma$) value represents the Fourier significance of a detected signal, expressed as an equivalent Gaussian significance level derived from the chi-squared ($\chi^2$) distribution \citep{1983ApJ...266..160L, 2001PhDT.......123R}.
During candidates sifting, an acceptable $\sigma$ threshold is selected to avoid a large number of false positives
and detect weak pulsars (e.g., $\sigma > 6$ in \citealt{2015ApJ...812...81L}; $\sigma > 4$ in \citealt{2024MNRAS.530.1436V}).
We used a detection threshold of $\sigma = 2$ in both the \texttt{accelsearch} routine and candidate sifting. This balanced sensitivity and false-positive rejection, allowing detection of faint pulsars near the survey sensitivity limit. M13I was detected at $3.7\,\sigma$ and $4.0\,\sigma$ in two epochs. A higher threshold would have excluded this weak but real pulsar. 

%\textcolor{blue}{\sout{Pulsar nulling and interstellar scintillation both}}
%\textcolor{blue}{\sout{cause apparent signal disappearances but have distinct origins.}}
%\textcolor{blue}{\sout{Nulling represents an intrinsic, broadband emission cessation,}}
%\textcolor{blue}{\sout{ while scintillation is a propagation effect that}} 
% \textcolor{blue}{\sout{induces frequency-dependent flux}}
%\textcolor{blue}{\sout{modulation.}}
 % (\citealt{1970Natur.228...42B, 1990ARA&A..28..561R}).}}

For M13G and M13H, their weak flux densities and sporadic detections suggest scintillation dominates their detectability. Both pulsars were observed during scintillation-enhanced states. Notably, M13H was detected in only 2 of 84 epochs. This implies that additional faint pulsars may remain undetected in our data but could become visible during future episodes of strong scintillation enhancement.

\section{Conclusions}\label{Conclusions} 

We conducted a deep acceleration search for pulsars in M13 using 84 FAST observations spanning from 2018 to 2025.
We had conducted a Fourier-domain acceleration search for all the observed full-length and segmented data, and also carried out a ``jerk'' search for the segmented data.

We discovered two new pulsars in binaries, namely J1641+3627G and H, or M13G and H.
We obtained a timing solution for M13G that spans the seven years of FAST data.
M13G is in a black window system
with an orbital period of 0.12 d and an extremely
low-mass companion, $M_{\rm c, med}$\,$\sim$\,9.9$\times 10^{-3}\,{\rm M}_\odot$.
M13G has the largest projected angular offset from M13's center.
%\textcolor{blue}{\sout{, providing further proof for the previous argument}} \textcolor{blue}
%{\sout{that the eccentricity of M13 binary systems decreases}}\textcolor{blue}{\sout{ with the projected distance from the cluster centre.}}
No eclipse was observed, which suggests this system could be in a low inclination angle system.

%$\mu {\rm Jy}$
M13H was discovered with a significant accelerated orbits. We failed to determine the timing solution of M13H due to a low detection rate of 2/84. The flux density of the two detections are 2.12 and 3.48\,$\mu {\rm Jy}$ respectively.

\begin{acknowledgements}
This work is supported by National Key Research and Development Programme, No. 2024YFA1611502 and National SKA Program of China (No. 2020SKA0120100),  the Basic Science Center Project of the National Nature Science Foundation of China (NSFC) under Grant Nos. 11703047, 11773041, U1931128, U2031119, 12003047, 12173052, 12373032 and 12173053.
This work is supported by the science and technology innovation Program of Hunan Province (No.2024JC0001).
Liyun Zhang has been
supported by the Science and Technology Program of Guizhou
Province under project No.QKHPTRC-ZDSYS[2023]003 and
QKHFQ[2023]003.
Lei Qian is supported by the Youth Innovation Promotion Association of CAS (id.~2018075, Y2022027), and the CAS ``Light of West China'' Program. 
%RPE is supported by the Chinese Academy of Sciences President's International Fellowship Initiative, Grant No. 2021FSM0004.
This work made use of the data from FAST (Five-hundred-meter Aperture Spherical radio Telescope) (https://cstr.cn/31116.02.FAST).  FAST is a Chinese national mega-science facility, operated by National Astronomical Observatories, Chinese Academy of Sciences.
We gratefully acknowledge the generous computational support provided by Guizhou Suanjia Computing Services Co., Ltd, Gui'an New Area Science and Technology Innovation Industries Development Limited Company, and Gui'an Supercomputing Center.
% This work is also supported by the International Centre of Supernovae, Yunnan Key Laboratory (No. 202302AN360001 and 202302AN36000104).
Finally, we thank the anonymous referee for helpful suggestions to bring clarity to the text.
\end{acknowledgements}

% Facility: FAST. Software:PRESTO (Ransom et al. 2002), DSPSR (van Straten \& Bailes 2011), TEMPO2 (Hobbs et al. 2006).

% \vspace{5mm}
% \facilities{HST(STIS), Swift(XRT and UVOT), AAVSO, CTIO:1.3m, CTIO:1.5m, CXO}


% \software{astropy \citep{2013A&A...558A..33A,2018AJ....156..123A},  
%          Cloudy \citep{2013RMxAA..49..137F}, 
%          Source Extractor \citep{1996A&AS..117..393B}
%          }



\bibliography{Refs_M13}{}
\bibliographystyle{aasjournal}

\appendix

\section{The profiles of M13G and M13H}

The pulse profile of M13G is complex, comprising both a main pulse and an interpulse. The main pulse consists of two distinct components, and an additional component exists between the main pulse and the interpulse (as shown in Figure~\ref{fig:timing}).
The pulse profile of M13H is single-peaked, but we are unable to assess the presence of potential substructure due to the limited signal-to-noise ratio.

\begin{figure*}[h!]
\begin{center}
\includegraphics[width=0.497\linewidth]{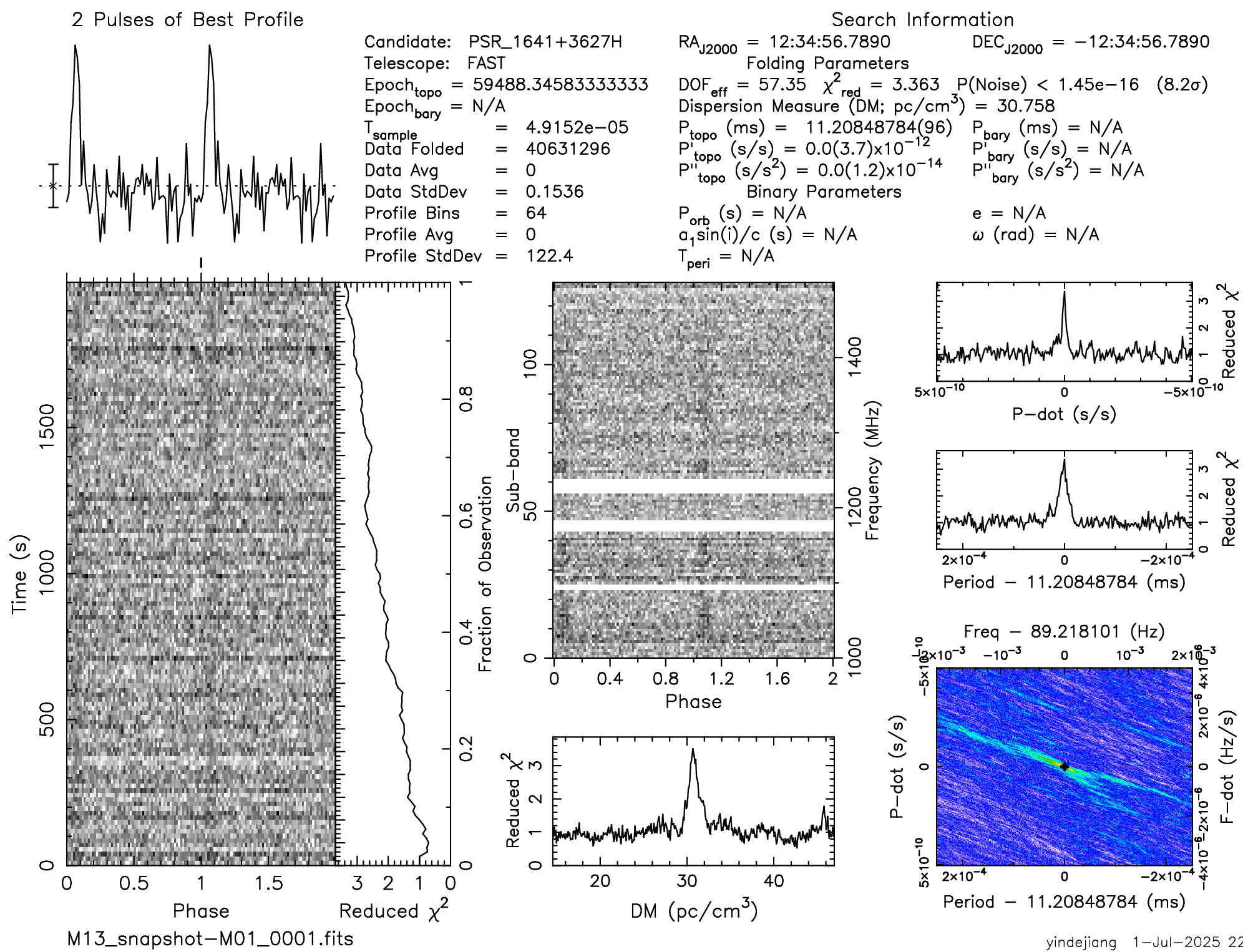}
\includegraphics[width=0.497\linewidth]{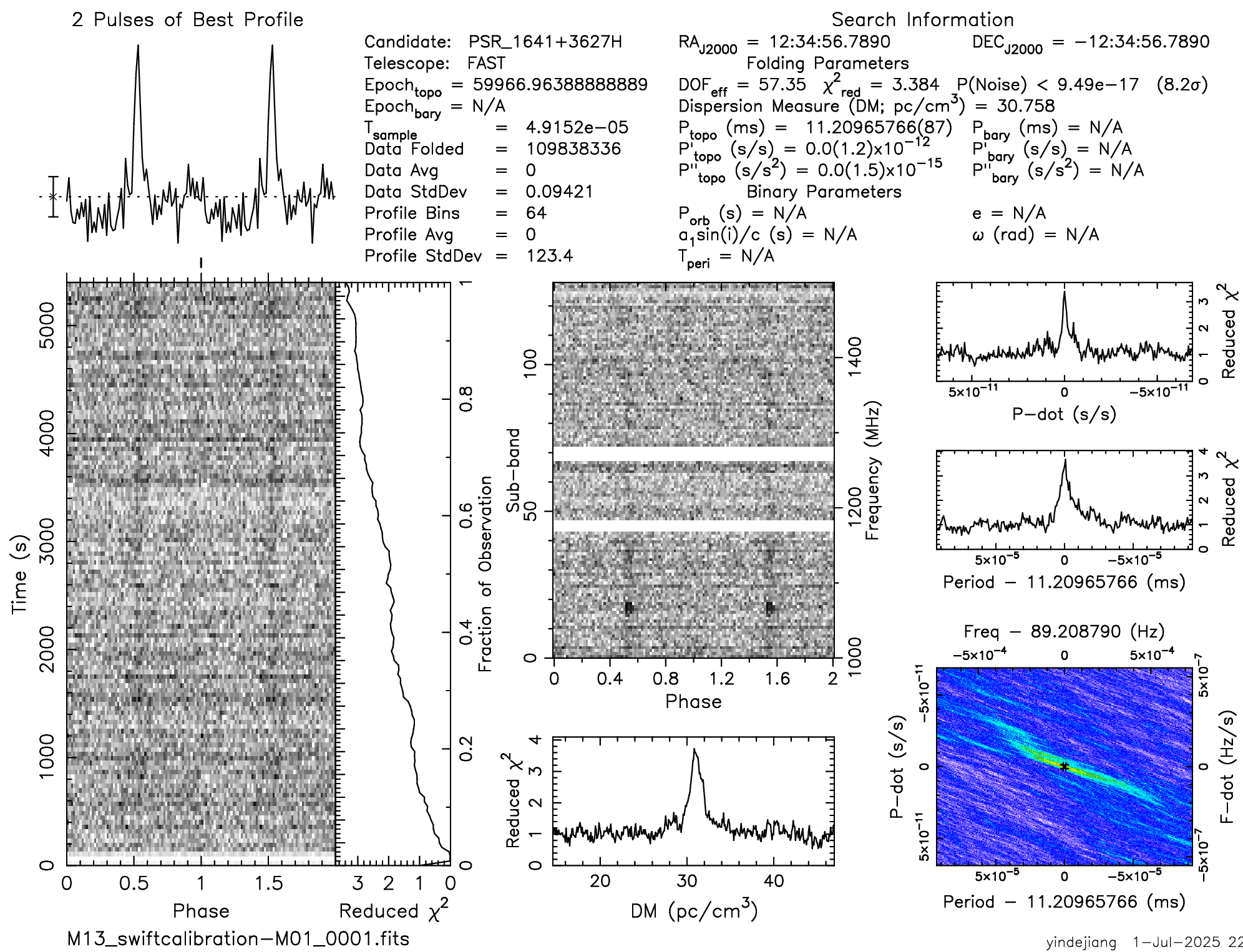}
\caption{
The M13H folding plots from the discovery and confirmation observations on MJD 59488 (left) and MJD 59966 (right).
%generated with the \texttt{prepfold} routine of \textsc{PRESTO}.
% The profiles show one full rotation of the pulsar with 64 phase bins, 256 sub-integrations and 128 frequency channels.
}
\label{fig:M13H}
\end{center}
\end{figure*}

To derive the orbital parameters, we made a temporary ephemeris of M13H  with ``BT'' model. 
PRESTO search results suggested M13H's orbital period was likely $\sim$0.5 days. 
So, we fixed the longitude of periastron $\omega$ =0, orbital eccentricity $e$= 0, 
and combined $P_b$, $T_{0\rm }$ and $x_{\rm p}$ to fit observed spin $P$ and $\dot{P}$ in the two datasets, with $P_b$ and $x_{\rm p}$ $<$ 1.
The \texttt{fitorb.py} returned many solutions due to the unconstrained data, and one of the optimal solutions was chosen to fold M13H shown in Figure \ref{fig:M13H}.
% A simple timing analysis was performed with this chose ephemeris using \textsc{TEMPO}.
The key parameters of the derived ephemeris for the folding were $f$ = 89.18380(2), $\dot{f}$ = 3.3(6)$\times 10^{-11}$, $P_b$ = 0.4299888(4) days, $x_{\rm p}$ = 2.949(1) lt-s, and $T_0$ = 59488.5263(3).

The diagnose plots shown here were generated by the \texttt{prepfold} routine of \textsc{PRESTO}.
The Fourier significance of a detected signal is derived by computing the reduced $\chi^{2}$ statistic for a model which assumes no pulsations.
The reduced $\chi^{2}$ is 3.36 with an effective degree of freedom of 57.35 in the left plot, while the corresponding parameters in the right plot are 3.38 and 57.35 (for details, see  \citealt{1983ApJ...266..160L, 2021ApJ...909...33B}).
The probability of the signal in the discovery plot which might be due to noise is $P_{\rm Noise}$\(< 1.45\times10^{-16}\), equivalent to a 8.2\,$\sigma$ Gaussian significance, while for the confirmation plot, this probability is $P_{\rm Noise}$\(< 9.49\times10^{-17}\), corresponding to a 8.2 $\sigma$ Gaussian significance.
The pulsar features (two vertical lines) can be easily identified in both phase versus time and phase versus observing frequency relation subplots of both diagnostic plots.

\section{Pulsar M13G Acceleration in the Cluster Field}\label{GC_field}

We noticed that M13G has negative observed spin period derivative ($\dot{P}_{\rm obs}$).
The intrinsic spin period derivative of pulsar should be positive (Radio pulsars are generally spin-powered).
Pulsars with negative $\dot{P}_{\rm obs}$ should be accelerating due to the intrinsic spin-down or other gravitational fields (e.g., \citealt{1993ASPC...50..141P}).
Since the updated parameters of previously known pulsars in M13 will be discussed separately (Wang et al., in preparation), here we focus on the acceleration of M13G within the cluster's gravitational potential.
The measured spin period derivative for a pulsar in GC is generally described by:
\begin{equation}
\left( \frac{\dot{P}}{P} \right)_{\rm obs} = \left( \frac{\dot{P}}{P} \right)_{\rm int} + \frac{\mu^2d}{c} + \frac{a_{\ell,\, \rm GC}}{c} + \frac{a_{g}}{c},
\end{equation}
where $\mu$ represents the total proper motion of the system, $d$ is the distance to the cluster ($6.16 \pm 0.44$ kpc, \citealt{2022ApJ...934..150L}), and the term $\mu^2d / c$ is known as so-called the Shklovskii effect (see \citealt{1970SvA....13..562S}). And $c$ denotes the speed of light, $a_{\ell,\, \rm GC}$ is the line-of-sight acceleration of the pulsar due to the gravitational field of the cluster, and $a_{g}$ is the acceleration of the center of mass of the GC in the potential of the Galaxy minus the Galactic acceleration of the solar system projected along the line of sight from the Earth to M13.

Since the proper motions of the M13G has not yet been well measured, but they are expected to be very similar to that of its host cluster, $-3.149 \pm 0.023 \, \rm mas ~ yr^{-1}$ and $-2.574 \pm 0.023 \, \rm mas ~ yr^{-1}$ (from $Gaia$ EDR3, \citealt{2021MNRAS.505.5978V}); thus $\mu^2d$ is $\sim$ $7.421 \times 10^{-11} \, \rm m \, s^{-2}$.
% Using the Galactic mass model derived by \citet{2017MNRAS.465...76M}, we find accelerations due to the Galactic potential of $ a \, \sim \, -0.165 \times 10^{-9}\, \rm m \, s^{-2}$ for M13.
$a_{g}$ can be estimated by (e.g., \citealt{1995ApJ...441..429N,2017ApJ...845..148P}): 
\begin{eqnarray}
a_{g} \cdot \vec{n} = -{\rm cos}(b)(\frac{\Theta_{0}^{2}}{R_{0}})({\rm cos}(l) + \frac{\beta}{{\rm sin}^{2}(l) + \beta^{2}}){\rm ~m~s^{-2}},
\end{eqnarray}
where $R_{0}$ is the distance of the Sun from the Galactic center (8.34$\pm$0.16~kpc) and $\Theta_{0}$ is the rotational speed of the Galaxy at the Sun's location (240$\pm$8~km~s$^{-1}$)  \citep{2014ApJ...793...51S}.
$\beta = (d/R_{0}){\rm cos}(b) - {\rm cos}(l)$, where $l$ and $b$ are the Galactic longitude and latitude of the GC, respectively. For M13, $l=59.01^{\circ}$ and $b=40.91^{\circ}$ giving $a_{g}\sim -9.702\times10^{-11}\,{\rm m\,s^{-2}}$ .
%  ##(5) Central velocity dispersion sig_v (km/s)
%  ## (7) Distance from Sun (kiloparsecs)
%  2022ApJ...934..150L
To model the acceleration along the line of sight caused by the field of the M13, $a_{\ell, \rm GC}$, an analytical model of the cluster developed by \citet{2005ApJ...621..959F} is utilized, which assumes the \citet{1962AJ.....67..471K} density profile
\begin{equation}
 a_{\ell,\, \rm GC}(x) = \frac{9 \sigma_{\mu,0}^2}{d \theta_c} \frac{\ell}{x^3}
 \left( \frac{x}{\sqrt{1+x^2}}  - \sinh^{-1}x \right),
 \end{equation}
where $x$ denotes the distance from the pulsar to the center of the GC divided by its core radius ($r_c=\theta_c d$),
$\sigma_{\mu,0}$ is the central stellar velocity dispersion of cluster (0.261$\pm0.006\, \rm mas ~ yr^{-1}$, \citealt{2022ApJ...934..150L}), and $\ell$ is the separation along the line-of-sight direction between the pulsar and the center of gravity of the cluster, also in units of $r_c$.
For each line of sight, we calculate the maximum and minimum values of $a_{\ell, \rm GC}(x)$, $a_{\ell, \rm max}$.
%In Figure~\ref{fig:acceleration}, the solid black lines represent the values of $a_{\ell, \rm max}$ as a function of angular distance from the center ($\theta_\perp$) of M13. The values of $a_{\ell, \rm max}$ for the lines of sight of each pulsar are listed in Table~\ref{table:acceleration}.
For M13G, an independent upper limit on the acceleration is determined from $\dot{P}_{\rm obs}$ as 
\begin{equation}
a_{\ell, \rm P, max} \, = \, c \frac{\dot{P}_{\rm obs}}{P} - \mu^2 d - a_g,
\end{equation}
which assumes $\dot{P}_\mathrm{int} = 0$.
The values of $a_{\ell, \rm P, max}$ for M13G is $\sim -0.44 \times 10^{-9} \rm m \, s^{-2}$ .
The described analytical model can account for the negative $\dot{P}_{\rm obs}$ of M13G which resides in the more distant half of cluster.

By assuming for one pulsar that the accelerations are the maximum and minimum allowed by the mass model of the cluster for their lines of sight ($\pm a_{\ell, \rm max}$), we can calculate lower and upper limits for the intrinsic spin period derivative $\dot{P}_{\rm int}$, which in turn allow us to determine minimum and maximum values for its surface magnetic field ($B_{\rm s} = 3.2\times 10^{19}(P\dot{P})^{1/2}\ G$) and characteristic age ($\tau_{\rm {c}} = \frac{P}{2\dot{P}}$). 
The $\dot{P}_{\rm int}$ of M13G was estimated to be $< 0.36 \times 10^{-20}\rm ~s~s^{-1}$, thus its $B_{\rm s} < 0.13 \times 10^9~\rm G$, $\tau_{\rm {c}} > 18.26~\rm  Gyr$.
The model-derived values of M13G are not very constrained, because the characteristic age has evolved beyond the cluster age of 12$\pm$0.38 Gyr \citep{2013ApJ...775..134V}.
For binaries, more precise estimated values will be obtained with the measurements of orbital period derivatives (e.g., \citealt{2017MNRAS.471..857F}).

% M13 age: 12 \pm 0.38 Gyr \citep{2013ApJ...775..134V}, why psrs were older than cluster ?

\begin{figure*}[htp!]
\begin{center}
\includegraphics[width=0.99\linewidth]{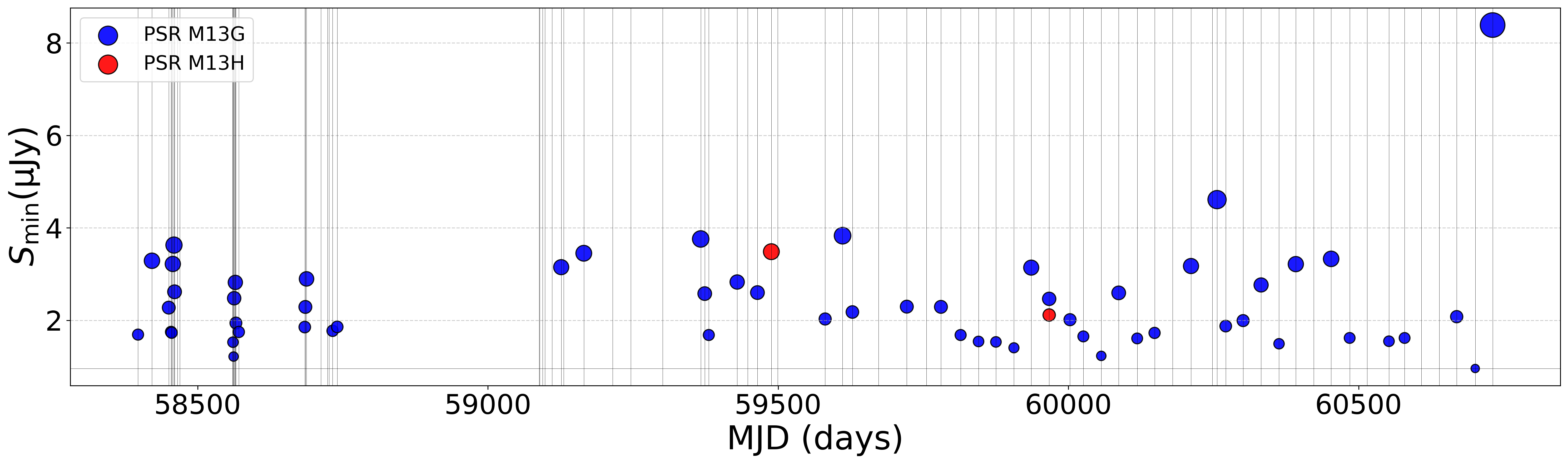}
\caption{
The variations of the the estimated flux (SNR $>$ 6
%\textcolor{red}{greater than 6? please check.} 
) for M13G and M13H are presented as a function of the Modified Julian Date (MJD). The gray vertical lines represent all observations used in this work.
The blue dot and the red dot represent M13G and M13H respectively, and the size of the dot is scaled by its estimated flux density.
}
\label{fig:snr}
\end{center}
\end{figure*}

% Binary Model                                                          \dotfill &   BT                                                                     \\
% Projected Semi-major Axis, $x_{\rm p}$ (lt-s)                         \dotfill &   1.6786(2)$\times 10^{-2}$                                              \\
% Orbital Eccentricity, $e$                                             \dotfill &   0.00$^b$                                                                     \\
% Longitude of Periastron, $\omega$ (deg)                               \dotfill &   0.00$^b$                                                                      \\
% Epoch of passage at Periastron, $T_0$ (MJD)                           \dotfill &   60256.359651(2)                                                        \\
% Orbital Period, $P_b$ (days)                                          \dotfill &   0.1205237186(4)     
% 

% \section{Appendix information}

% Appendices can be broken into separate sections just like in the main text.
%The only difference is that each appendix section is indexed by a letter
%(A, B, C, etc.) instead of a number.  Likewise numbered equations have
%the section letter appended.  Here is an equation as an example.

%%%%% TEMPO
\begin{table}[htbp!]
% \begin{table}[H]
\begin{center}{\scriptsize
\setlength{\tabcolsep}{4pt}
\renewcommand{\arraystretch}{1.3}
\caption{The timing solution of M13G. 
}\label{table:timing}
\begin{tabular}{lc}
\hline
Pulsar  &   J1641+3627G                                                            \\
\hline\hline
Right Ascension, $\alpha$ (J2000)                                     \dotfill &  16:41:42.9541(1)                                                     \\
Declination, $\delta$ (J2000)                                         \dotfill &  +36:29:21.717(2)                                                    \\
Spin Frequency, $f$ (s$^{-1}$)                                        \dotfill &   231.29059200665(1)                                                     \\
1st Spin Frequency derivative, $\dot{f}$ (s$^{-2}$)                \dotfill &   3.548(2)$\times 10^{-16}$                                              \\
Reference Epoch (MJD)                                                 \dotfill &   60256.228538                                                           \\
Start of Timing Data (MJD)                                            \dotfill &   58397.302                                                              \\
End of Timing Data (MJD)                                              \dotfill &   60731.003                                                              \\
Dispersion Measure, DM (pc cm$^{-3}$)                                 \dotfill &   30.759(4)                                                              \\
Solar System Ephemeris                                                \dotfill &   DE438                                                                  \\
Terrestrial Time Standard                                             \dotfill &   UTC(NIST)                                                              \\
Time Units                                                            \dotfill &   TDB                                                                    \\
Number of TOAs                                                        \dotfill &   249                                                                    \\
Residuals RMS ($\mu$s)                                                \dotfill &   24.27                                                                  \\
\hline
\multicolumn{2}{c}{Binary Parameters}  \\
\hline\hline
Binary Model                                                          \dotfill &   BT                                                                     \\
Projected Semi-major Axis, $x_{\rm p}$ (lt-s)                         \dotfill &   1.6787(2)$\times 10^{-2}$                                              \\
Orbital Eccentricity, $e$                                             \dotfill &   0.0$^a$                                                                      \\
Longitude of Periastron, $\omega$ (deg)                               \dotfill &   0.0$^a$                                                                      \\
Epoch of passage at Periastron, $T_0$ (MJD)                           \dotfill &   60256.359652(2)                                                        \\
Orbital Period, $P_b$ (days)                                          \dotfill &   0.1205237188(4)                                                        \\
\hline
\multicolumn{2}{c}{Derived Parameters}  \\
\hline\hline
Spin Period, $P$ (s)                                                  \dotfill &   4.3235653959122(2)$\times 10^{-3}$                                     \\
1st Spin Period derivative, $\dot{P}$ (s s$^{-1}$)                    \dotfill &   $-$6.633(3)$\times 10^{-21}$                                           \\
Mass Function, $f(M_{\rm p})$ (${ M}_\odot$)                       \dotfill &   3.4967$\times 10^{-7}$                                              \\
Minimum companion mass, $M_{\rm c, min}$ (${ M}_\odot$)            \dotfill &   0.0086$^b$                                                   \\
Median companion mass, $M_{\rm c, med}$ (${ M}_\odot$)             \dotfill &   0.0099$^b$                                                    \\
Offset from GC center in $\alpha$, $\theta_\alpha$ (arcmin)           \dotfill &  $-$0.3444$^c$                                                                   \\
Offset from GC center in $\delta$, $\theta_\delta$ (arcmin)           \dotfill &   1.7702$^c$                                                               \\
Total offset from GC center, $\theta_\perp$ (arcmin)                  \dotfill &   1.8034$^c$                                                                     \\
Projected distance from GC center, $r_\perp$ (pc)                     \dotfill &   2.9087$^c$                                                               \\
Projected distance from GC center, $r_\perp$ (core radii)             \dotfill &   3.2314$^c$                                                                \\
\hline
\end{tabular} }
%\tablefoot{% The DE438 solar system Ephemeris, the Barycentric Dynamic Time time units (TDB) and Terrestrial Time Standard UTC(NIST) time standard are used.
% $^a$The proper motion of M13G should be consistent with its host cluster. The fitting error of proper motion in $\delta$ was currently more than 3-$\sigma$ and therefore the proper motion of the M13 from \citet{2022ApJ...934..150L} was fixed for fitting. 
% $^a$The fitting error of proper motion in $\delta$ was currently more than 3$\sigma$ and therefore the average proper motion of the M13, $-3.149 \pm 0.023 \, \rm mas ~ yr^{-1}$ and $-2.574 \pm 0.023 \, \rm mas ~ yr^{-1}$ for $\alpha$ and $\delta$, respectively (from $Gaia$ EDR3 \citealt{2021MNRAS.505.5978V}), was fixed for fitting. 
% $^a$The eccentricity and the longitude of periastron for this pulsar are set quantities. $^b$The companion mass is derived from radio timing and assume a pulsar mass of $1.35 \, \rm M_{\odot}$. The minimum and median masses assume an orbital inclination angle of $i\, = \,90^\circ$ and $i\, = \,60^\circ$, respectively. $^c$The M13G’s position offset is derived from the cluster center ($\alpha_{\rm J2000} = 16^{\rm h}41^{\rm m}41.24^{\rm s}$, $\delta_{\rm J2000} = +36^{\rm \circ}27^{\rm '}35.5^{\rm ''}$) for a cluster distance of 6.16 kpc \citep{2022ApJ...934..150L}.}% \textcolor{red}{give citation}}      
    \begin{itemize}
        \item[]
    \textbf{Notes:} $^a$The eccentricity and the longitude of periastron for this pulsar are set quantities. $^b$The companion mass is derived from radio timing and assume a pulsar mass of $1.35 \,  M_{\odot}$. The minimum and median masses assume an orbital inclination angle of $i\, = \,90^\circ$ and $i\, = \,60^\circ$, respectively. $^c$The M13G’s position offset is derived from the cluster center ($\alpha_{\rm J2000} = 16^{\rm h}41^{\rm m}41.24^{\rm s}$, $\delta_{\rm J2000} = +36^{\rm \circ}27^{\rm '}35.5^{\rm ''}$) for a cluster distance of 6.16 kpc \citep{2022ApJ...934..150L}.
    \end{itemize}
\end{center} 
\end{table}
%% Mp = 1.35 Msun 
% Mass Function:          3.496672e-07 Msun
% Minimum Mc:             8.642148e-03 Msun
% Median Mc:              9.985670e-03 Msun
% Distance in RA:             -0.344362 arcmin
% Distance in DEC:             1.770190 arcmin
% Total Distance:              1.803386 arcmin
% Total Distance:              2.908688 core radii
% Total Distance:              3.231437 pc (for an assumed distance of 6.160 kpc)
%Weighted RMS residual: pre-fit    24.270 us. Predicted post-fit    24.270 us.
% Chisqr/nfree:    362.60/  241 =     1.504571122   pre/post:   1.00   Wmax:   10.2

\end{document}